\documentclass[acmsmall]{acmart}

\AtBeginDocument{%
  }

\acmJournal{JACM}
\acmVolume{37}
\acmNumber{4}
\acmArticle{111}
\acmMonth{8}

\usepackage{subfigure}
\usepackage{enumerate}
\usepackage{adjustbox}

\begin{document}

%%
%% The "title" command has an optional parameter,
%% allowing the author to define a "short title" to be used in page headers.
\title{A Survey of Adversarial CAPTCHAs on its History, Classification and Generation}

%%
%% The "author" command and its associated commands are used to define
%% the authors and their affiliations.
%% Of note is the shared affiliation of the first two authors, and the
%% "authornote" and "authornotemark" commands
%% used to denote shared contribution to the research.

\author{Zisheng Xu}
\affiliation{%
  \institution{Shenzhen University}
  \streetaddress{3688 Nanhai Avenue}
  \city{Shenzhen}
  \country{China}}
\email{xuzisheng2021@email.szu.edu.cn}

\author{Qiao Yan}
\authornote{Corresponding author}
\affiliation{%
  \institution{Shenzhen University}
  \city{Shenzhen}
  \country{China}}
\email{yanq@szu.edu.cn}

\author{F. Richard Yu}
\affiliation{%
  \institution{Shenzhen University}
  \city{Shenzhen}
  \country{China}}
\email{richard.yu@carleton.ca}

\author{Victor C. M. Leung}
\affiliation{%
  \institution{Shenzhen University}
  \city{Shenzhen}
  \country{China}}
\email{vleung@ieee.org}

%%
%% By default, the full list of authors will be used in the page
%% headers. Often, this list is too long, and will overlap
%% other information printed in the page headers. This command allows
%% the author to define a more concise list
%% of authors' names for this purpose.
\renewcommand{\shortauthors}{Xu et al.}

%%
%% The abstract is a short summary of the work to be presented in the
%% article.
\begin{abstract}
Completely Automated Public Turing test to tell Computers and Humans Apart, short for CAPTCHA, is an essential and relatively easy way to defend against malicious attacks implemented by bots. The security and usability trade-off limits the use of massive geometric transformations to interfere deep model recognition and deep models even outperformed humans in complex CAPTCHAs. The discovery of adversarial examples provides an ideal solution to the security and usability trade-off by integrating adversarial examples and CAPTCHAs to generate adversarial CAPTCHAs that can fool the deep models. In this paper, we extend the definition of adversarial CAPTCHAs and propose a classification method for adversarial CAPTCHAs. Then we systematically review some commonly used methods to generate adversarial examples and methods that are successfully used to generate adversarial CAPTCHAs. Also, we analyze some defense methods that can be used to defend adversarial CAPTCHAs, indicating potential threats to adversarial CAPTCHAs. Finally, we discuss some possible future research directions for adversarial CAPTCHAs at the end of this paper.
\end{abstract}

%%
%% The code below is generated by the tool at http://dl.acm.org/ccs.cfm.
%% Please copy and paste the code instead of the example below.
%%
\begin{CCSXML}
<ccs2012>
   <concept>
       <concept_id>10010147.10010178</concept_id>
       <concept_desc>Computing methodologies~Artificial intelligence</concept_desc>
       <concept_significance>500</concept_significance>
       </concept>
 </ccs2012>
\end{CCSXML}

\ccsdesc[500]{Computing methodologies~Artificial intelligence}

%%
%% Keywords. The author(s) should pick words that accurately describe
%% the work being presented. Separate the keywords with commas.
\keywords{CAPTCHA, Deep learning, Adversarial examples, Adversarial CAPTCHAs}

% \received{20 February 2007}
% \received[revised]{12 March 2009}
% \received[accepted]{5 June 2009}

%%
%% This command processes the author and affiliation and title
%% information and builds the first part of the formatted document.
\maketitle

\section{Introduction}
Completely Automated Public Turing test to tell Human and Computer Apart, short for CAPTCHA \cite{VonAhn2003}, or Human Interaction Proofs (HIP) \cite{Baird2002}, is a test to distinguish humans and computers. The basic process of CAPTCHA verification is that the CAPTCHA system provides a puzzle for a user to solve. The puzzle should be only solvable by humans. Once the user submitted the answer, The CAPTCHA system checks the correctness of the answer and allows the user to pass the test if the answer is correct. This verification process can be seen as a reverse Turing test, this concept was first introduced by Naor \cite{Naor1996}. Turing test was proposed by Alan Turing \cite{Turing2009}, the test is conducted and judged by humans, while in the scenario of CAPTCHA verification, the test is conducted and judged by computers. Over decades of development, various types of CAPTCHAs were invented.  Common types of CAPTCHAs are as follows:
\begin{enumerate}[a)]
    \item{Text-based CAPTCHAs require users to type the characters within the presented image. Examples: \cite{VonAhn2008,Mori2003,Chew2003, Yan2008,Kim2019,Nguyen2012,Rusu2004,Chellapilla2005,Thomas2010,Yalamanchili2011,Imsamai2010,Rusu2009,Bursztein2014,Parvez2020,Yu2016,Saini2013}}
    \item{Image-based CAPTCHAs require users to select the objects hinted in the prompt. Examples: \cite{Elson2007,Vikram2011,DSouza2012,GoswamiF2014,GoswamiFR2014}}
    \item{Audio-based CAPTCHAs require users to transcribe the content in the audio. Examples: \cite{Kochanski2002,Markkola2008,Meutzner2015,VonAhn2008}}
    \item{Drag-based CAPTCHAs require users to drag certain components in the CAPTCHA to fill the pattern. Examples: \cite{Acien2020,Mohamed2016,Zhao2018}}
    \item{Video-based CAPTCHAs require users to answer the question in the prompt based on users' comprehension of the video. Examples: \cite{Kluever2009,Shirali2008,Rao2016,Chow2011,Cui2010}}
\end{enumerate}

CAPTCHA plays an important role in the field of cybersecurity. The first form of CAPTCHA could date back to 1999 when a website named Alta Vista wanted to prevent database modification by bots. Except for this application, in the domain of cybersecurity and communications, CAPTCHA can prevent denial of service (DoS) attack. DoS attack is an attack that restricts users' system access by fully occupying the target server \cite{Carl2006}. When a DoS attack is triggered, a single bot or system sent excessive requests to the target server, and the server responds to the request and allocates resources for those requests. However, the adversaries keep occupying these resources and not releasing them, draining out the resources of the target server and the target server cannot allocate any resources. Similar to DoS attack, a DDoS (Distributed Denial of Service) attack utilizes multiple bots or systems to send requests \cite{Yan2016}. When the target server is attacked via DoS or DDoS attack, it can present a CAPTCHA test to the sender of the requests. If the request sender failed to pass the CAPTCHA test, the target server recycles the allocated resources and thus still is able to respond to normal requests. Similar to DoS attack, dictionary attack attempts to crack the password by permuting all the possible combinations of the characters and constantly querying the target system \cite{Bosnjak2018}. The attacked target system can block the dictionary attack by applying CAPTCHA tests. The bot or computer that is used to implement dictionary attack cannot solve the CAPTCHA test and thus the target system block the query process in the dictionary attack \cite{Chakrabarti2007,Pinkas2002}.

A successful CAPTCHA should satisfy two conditions, security and usability. The test presented by the CAPTCHA system should not be passed by computers (security), while humans can pass the CAPTCHA test effortlessly, regardless of age, gender, education level, or language (usability). However, the development of computer science makes computers intelligent enough to solve some CAPTCHA tests. An adversary may attempt to bypass the CAPTCHA system. One way to do so is traditional Optical Character Recognition, short for OCR. However, OCR might fail when massive distortion is applied to the CAPTCHA. Another way to crack a CAPTCHA system is deep model recognition. To date, deep models can be used to recognize images, audio, videos, or perform other tasks. Thus, it is possible for an adversary to bypass CAPTCHA systems automatically by recognizing them, and relative works have been proposed \cite{Chellapilla2004,Gao2016,Sivakorn2016}. One way to counter this situation is to apply more geometric transformations. However, this countermeasure has its disadvantage, that is, excessive transformations could fail human recognition, thus sabotaging the usability requirement for CAPTCHAs. Another way to bypass CAPTCHA systems is by utilizing the internal weakness in the CAPTCHA systems \cite{Fritsch2010,Hernandez2010}.

The security and usability trade-off of applying geometric transformations to CAPTCHAs haunted developers for years until the discovery of adversarial examples. Szegedy et al. discovered that deep learning models are vulnerable to some carefully crafted and imperceptible perturbations. When the perturbations are applied to the correctly classified original image as a new input to the deep models, the deep models misclassify it with high confidence. Such input is called adversarial examples. The ability to make deep models misclassify while remaining imperceptible for humans makes adversarial examples a perfect tool to solve the security and usability trade-off. CAPTCHAs now cannot be solved by computers without applying excessive geometric transformations. Although adversarial examples were first discovered in 2014 and intrigued many researchers, currently there is no such work to systematically summarize the applications of adversarial examples in CAPTCHAs. Therefore, this paper intends to provide a systematic review of adversarial CAPTCHAs to achieve a better understanding of the previous works, and thus inspires future works of adversarial CAPTCHAs.

The main contributions of this paper are listed as follows:
\begin{enumerate}
    \item{To our knowledge, this is the first systematic survey in the domain of adversarial CAPTCHAs.}
    \item{We extended and re-formalized the definition of adversarial CAPTCHAs,  briefly reviewed the history of CAPTCHA development, and analyzed the problem that existed in CAPTCHAs.}
    \item{We proposed a classification method for adversarial CAPTCHA in the perspective of semantic information, then categorized the related works of adversarial CAPTCHAs after reviewing them.}
    \item{We presented some challenges for adversarial CAPTCHAs and discussed some directions for future research on adversarial CAPTCHAs}
\end{enumerate}

\begin{figure*}
\centering
\includegraphics[width=\linewidth]{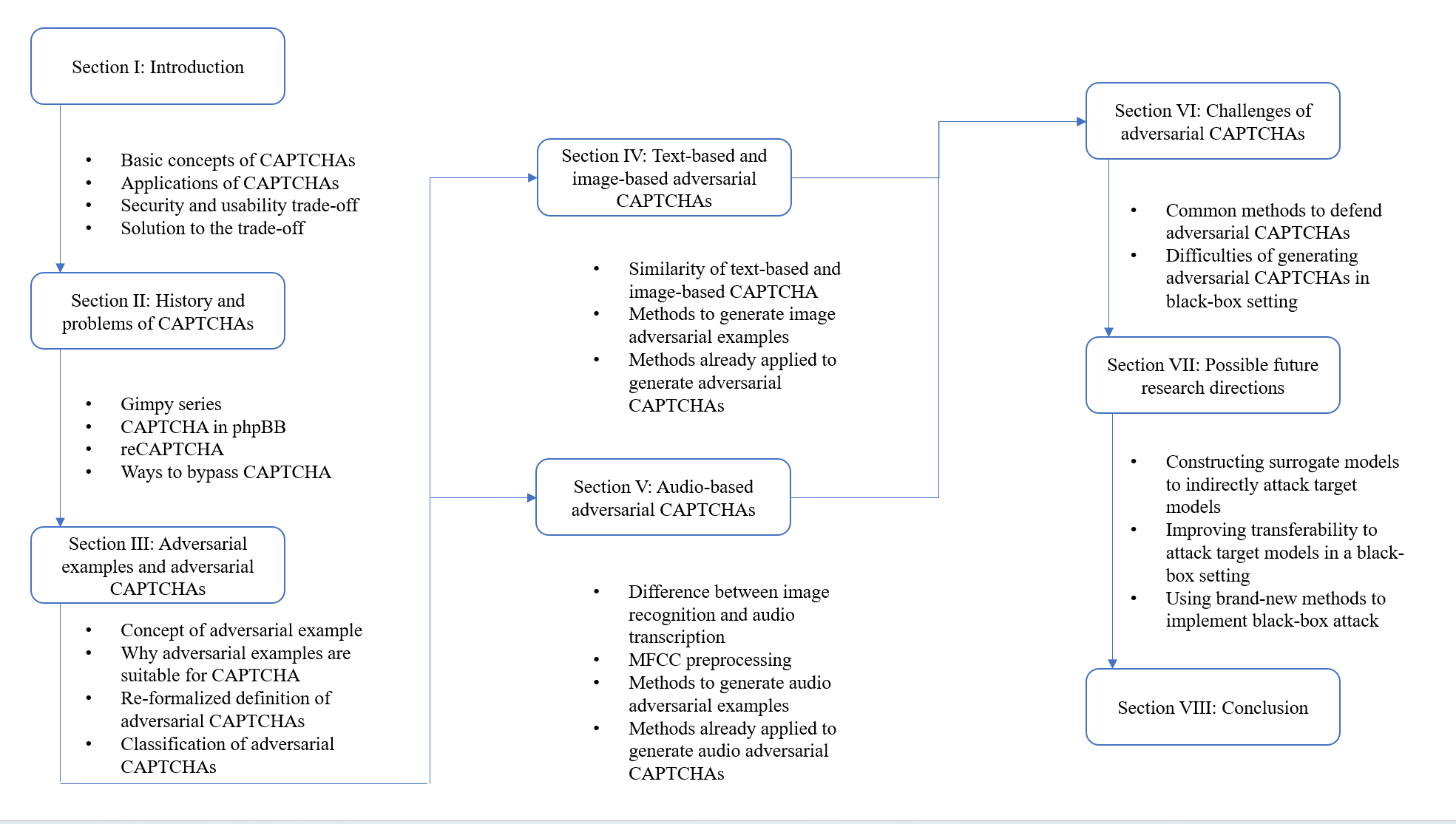}
\caption{Overall structure of this paper}
\label{fig_1}
\end{figure*}

The remaining structures of this paper are shown in Figure \ref{fig_1} and listed as follows:
\begin{enumerate}
    \item{Section II introduced the history of CAPTCHA and some classical CAPTCHA schemes. Then we listed the disadvantage of the current CAPTCHA scheme}
    \item{Section III gave a brief introduction to adversarial examples and stated the need for adversarial CAPTCHA. We explained the definition of adversarial CAPTCHAs and proposed a new classification method for adversarial CAPTCHAs}
    \item{Section IV introduced the verification process of text-based and image-based CAPTCHA and some generation methods for image adversarial examples, also with some examples of adversarial CAPTCHA schemes}
    \item{Section V introduced the verification process of audio-based CAPTCHA and some generation methods for audio adversarial examples}
    \item{Section VI listed some current challenges and limitations for adversarial examples}
    \item{Section VII presented some possible future research directions for adversarial CAPTCHAs}
\end{enumerate}
%% The Appendices part is started with the command \appendix;
%% appendix sections are then done as normal sections
%% \appendix

% -----------------------------History and problems of CAPTCHAs--------------------------------------

\section{History and problems of CAPTCHAs}
\begin{figure*}
\centering
\includegraphics[width=\linewidth]{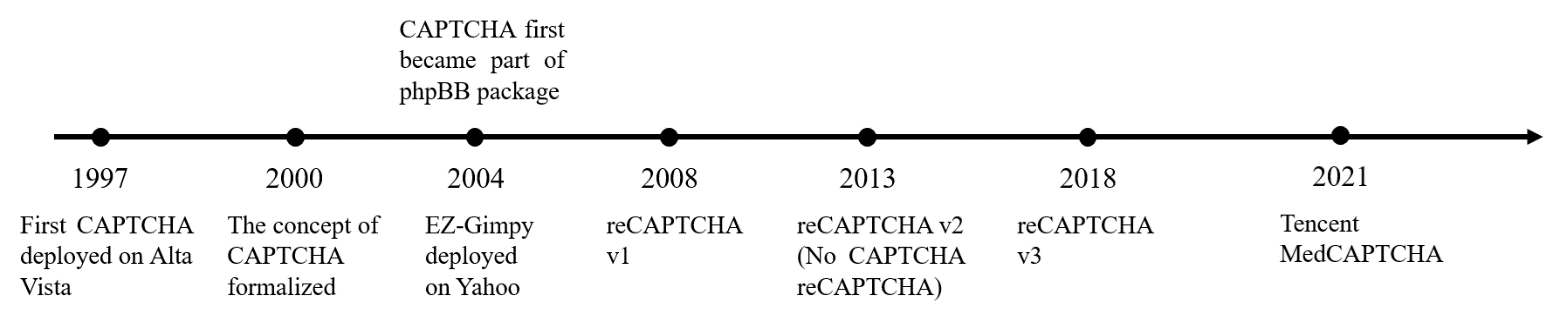}
\caption{Timeline of CAPTCHA development}
\label{fig_2}
\end{figure*}

The form of CAPTCHA first appeared in 1997, when the search engine Alta-Vista wanted to find a way to prevent bots or automated computer programs from adding spam and malicious URLs (Uniform Resource Locator) to their database \cite{Brodic2018}. Later in 2000, Louis Von Ahn et al. formalized the term CAPTCHA to the term Complete Automated Public Turing test to tell Computers and Humans Apart \cite{VonAhn2003}. CAPTCHA is a test that needs to be passable by any human, regardless of age, gender, education level, or language. Years later, different CAPTCHAs were developed, such as  Gimpy \cite{Mori2003}, reCAPTCHA \cite{VonAhn2008}, and Tencent MedCAPTCHA \cite{SZU2021}. Tencent MedCAPTCHA first presents users with an example to annotate the corresponding body tissue and requires users to annotate the same tissue in a new image. Body tissues vary in different people, different ages and different disease progress, making DNN models hard to accurately segment the body tissues, thus being able to distinguish humans and computers. Meanwhile, users help medical institutes annotate medical images while solving the MedCAPTCHA. Here we listed some classic CAPTCHA systems in history in Figure \ref{fig_2}.

\subsection{Gimpy, EZ-Gimpy, and Gimpy-r}
Gimpy is one of the first CAPTCHAs proposed by Carnegie Mellon University \cite{Mori2003}. Gimpy selects several English words randomly and distorts them with non-linear deformation and overlapping. Recognizing multiple words at one time increased the verification time and decreased the overall user experience, thus it is not suitable for commercial use. EZ-Gimpy (easy Gimpy) is a simplified version of Gimpy that used only one word for CAPTCHA verification. It was first employed by Yahoo email and Chatter to avoid malicious registration by robots. Unlike Gimpy and EZ-Gimpy, which takes word or words from a small dictionary, Gimpy-r picks random letters instead of words. By doing so, Gimpy-r has the flexibility to generate diverse CAPTCHAs. Besides, the letters in Gimpy-r are from different fonts.

\begin{figure}
\centering
\includegraphics[width=0.3\linewidth]{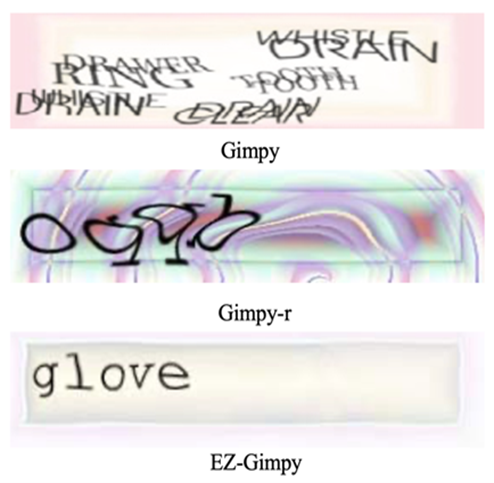}
\caption{Examples of Gimpy, EZ-Gimpy, and Gimpy-r}
\label{fig_3}
\end{figure}

\subsection{CAPTCHAs in phpBB}
CAPTCHA first became part of the phpBB package with version 2.0.10 in order to stop bots from maliciously registering accounts \cite{Kellanved2008}. The original CAPTCHAs of phpBB are generated by adding random noise to the background of the characters. It is simple to break through in the present day but it did the job back then. To improve the security of the CAPTCHA, a new version is introduced. In the new CAPTCHA, geometric transformations like distortion and rotation are applied to the characters. The characters are formed with hollow blocks and the size of the characters are different. To date, the latest version of CAPTCHA works reasonably well though there are lots of ways to break through the CAPTCHA system.

\begin{figure}
\centering
\subfigure[]{\label{fig_4a}\includegraphics[width=0.5\linewidth]{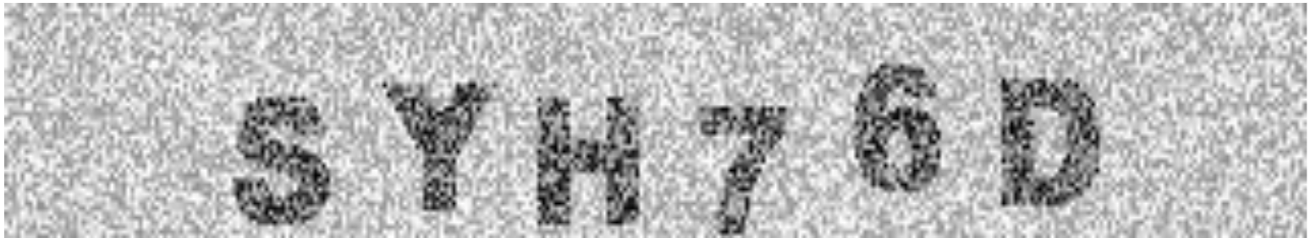}}
\subfigure[]{\label{fig_4b}\includegraphics[width=0.5\linewidth]{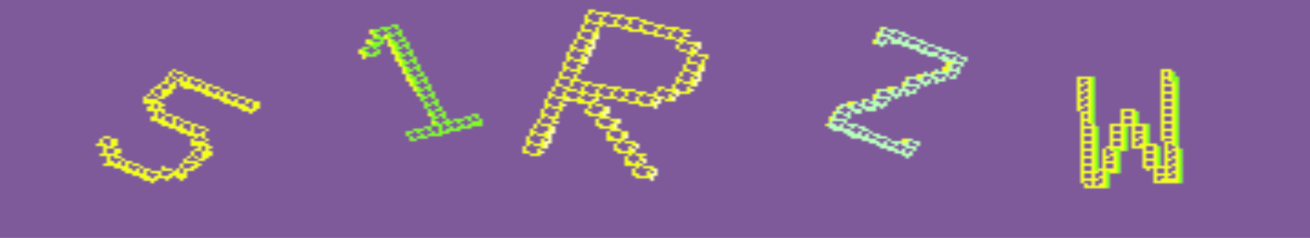}}
\caption{(a) The original version of CAPTCHA used in phpBB. (b) The new version of CAPTCHA used in phpBB.}
\label{fig_4}
\end{figure}

\subsection{reCAPTCHA}
A novel CAPTCHA scheme named reCAPTCHA was introduced in 2008 \cite{VonAhn2008}. reCAPTCHA v1 is text-based CAPTCHA, it presents two words to the user, one unknown word from the scanned text that OCR cannot recognize and one control word that the system knows the answer. The user needs to type two words correctly. The reCAPTCHA system assumes that the unknown word is also correct if the user types the control word correctly. To avoid human error and malicious attacks, suspicious words will be sent to multiple users. If multiple users give the same answer, the system assumes all users give the correct answer. Because the unknown words are scanned from ancient books, reCAPTCHA v1 is digitalizing books using the correct answers provided by the users, while maintaining the purpose of verification. Given the fact that reCAPTCHA v1 can be cracked, Google suspended it in 2018.

Later in 2013, Google presents reCAPTCHA v2, which is a “no CAPTCHA reCAPTCHA”. reCAPTCHA v2 first presents an “I’m not a robot” checkbox to users. By clicking the checkbox, the reCAPTCHA system runs a background check that analyses search history and cookies. If the CAPTCHA system cannot be sure about the identity of the user, then an image-based CAPTCHA is presented. It splits the image into a 3x3 or 4x4 grid and the users need to select grids that contain the specified object. Similar to reCAPTCHA v1, suspicious answers will be sent to multiple users, and users’ answers provide annotation to the images.

As of the latest version of reCAPTCHA, reCAPTCHA v3 completely gets rid of the need to input anything. The moment when the user visits the website deployed reCAPTCHA v3, the reCAPTCHA system automatically monitors the user’s behavior like the track of mouse and keyboard inputs. Similar to reCAPTCHA v2, reCAPTCHA v3 analyzes these behaviors and gives the final results.

For visual-impaired users, reCAPTCHA also provides audio CAPTCHA, we will discuss audio CAPTCHA in section VI.

\begin{figure}[!t]
\centering
\subfigure[]{\label{fig_5a}\includegraphics[]{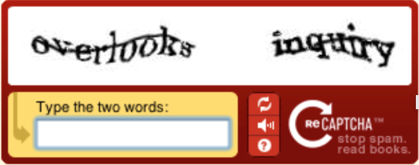}}
\subfigure[]{\label{fig_5b}\includegraphics[]{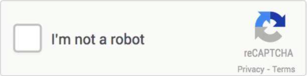}}
\subfigure[]{\label{fig_5c}\includegraphics[]{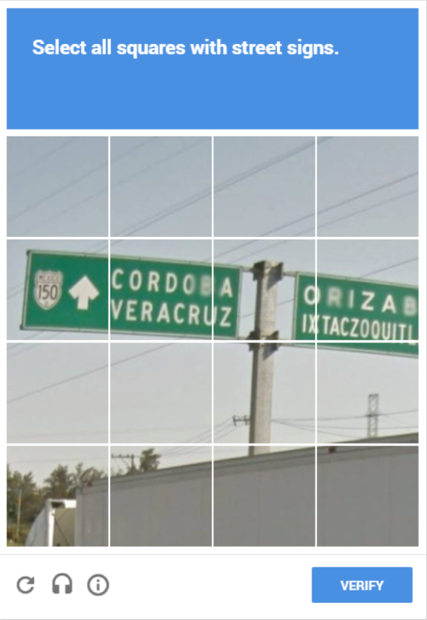}}
\caption{(a) Example of reCAPTCHA v1. (b) “I’m not a robot” checkbox. (c) The secondary check was presented when automatic verification failed in reCAPTCHA v2.}
\label{fig_5}
\end{figure}

Currently, there are several ways to bypass CAPTCHA schemes. Initially, traditional OCR was applied to recognize the characters in the CAPTCHA. To avoid OCR, geometric transformations were used in the CAPTCHA to make the characters distort, overlap, and hard to be recognized by OCR. Because humans are good at it, regardless of the angle, distortion, or light condition, while computers of the era are bad at it. The invention and evolution of the Graphics Processing Unit (GPU) and Tensor Processing Unit (TPU) have greatly expanded the learning capacity of computers. With the improvement of the accuracy of deep model recognition, CAPTCHAs can be recognized by deep models automatically and sometimes even outperform human recognition.

Another approach used a different way to validate real users, instead of recognizing the CAPTCHAs automatically. Apple added a new feature called automatic verification in iOS 16, iPad OS 16, and macOS Ventura \cite{Apple2022}. The key idea of automatic verification is that users already performed actions that are hard for bots to imitate before loading a web page, so validating users’ identities with CAPTCHAs is unnecessary. This new feature utilized Private Access Tokens to avoid CAPTCHAs. Private Access Token is based on Private Pass Protocol \cite{Davidson2018}, which is currently being standardized by IETF. When a user accesses a web server over HTTP (Hyper Text Transfer Protocol), the server sends back a challenge using the PrivateToken authentication scheme. This challenge requires the user to provide a token signed by the token issuer trusted by the server. In order to fetch the token, the user contacts the iCloud attester and sends a token request. The attester performs device attestation and verifies if the user's iCloud account is in good standing. If the user can be validated, the iCloud attester sends a request for a new token to the token issuer. Since the user is trusted by the attester, the token issuer signs the token. Finally, the user presents the signed token to the server, and the server checks if the token is valid. The whole process is imperceptible to the users, which remains a good user experience.

%% \section{}
%% \label{}

% ----------------------------Adversarial examples and adversarial CAPTCHAs----------------------------

\section{Adversarial examples and adversarial CAPTCHAs}
Although deep models show great effectiveness in various domains like image recognition, and autonomous driving, research shows that they are vulnerable to carefully crafted samples. Szegedy et al. discovered adversarial examples in 2013 \cite{Szegedy2013}. Suppose an image can be classified correctly by a model. By adding carefully treated, quasi-imperceptible perturbations to the image, the image became an adversarial example and makes the model output an incorrect classification result. Figure \ref{fig_4} gives a demo of an adversarial example, where the original image is correctly labeled as a panda, but when added carefully crafted perturbations, the model misclassified the image to a gibbon.

\begin{figure}[!t]
\centering
\includegraphics[width=0.8\linewidth]{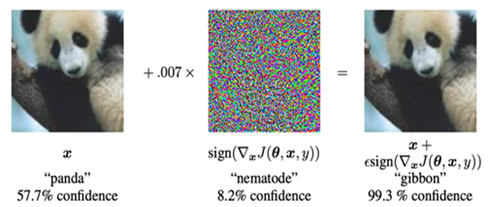}
\caption{Adversarial example generated by FGSM by Goodfellow et al. \cite{Goodfellow2014}}
\label{fig_6}
\end{figure}

Throughout the verification process of CAPTCHAs, one can obtain an image library of distorted characters, and use these data to train a machine learning or deep learning model to recognize new CAPTCHAs. According to a security blog posted by Google in 2014 \cite{Shet2014}, they launched an experiment to make humans and AI recognize the same most distorted texts. The human could solve these distorted text puzzles with 33\% accuracy, while the AI achieved 99.8\% accuracy, and outperformed humans dramatically. In this case, CAPTCHAs with the most distorted characters are useless since they cannot distinguish humans and computers. Adding too much distortion isn’t a good way to set a roadblock for CAPTCHA recognition, that’s why adversarial CAPTCHAs are introduced. 

Terada et al. \cite{Terada2022} proposed a definition of adversarial CAPTCHA, that is, ``a method that adds resistance against attacks that use machine learning classifiers to attack a CAPTCHA by applying adversarial example techniques to CAPTCHAs that are considered able to distinguish whether the operator is a human or a computing machine.'' Because the perturbations added are small, human beings usually cannot distinguish between benign and adversarial examples. Making full use of this property, one can apply adversarial examples to CAPTCHAs, leaving good usability while fooling deep models simultaneously. However, adversarial CAPTCHA under this definition limits its adversarial perturbations to be imperceptible. Except for adversarial examples, other approaches such as adversarial patches \cite{Brown2017} and unrecognizable images \cite{Nguyen2015} can also fool DNN models and be used to generate adversarial CAPTCHAs based on the difference between humans' and computers' prediction, while the modification to the benign images is perceptible to humans. Hence, we extend and re-formalize the definition of adversarial CAPTCHA as follows: 

\begin{definition}[Adversarial CAPTCHA]
Adversarial CAPTCHA is a method that defends against attacks that use machine learning models to recognize CAPTCHAs automatically, which is done by applying any techniques that can fool the machine learning models to CAPTCHAs, in order to distinguish humans and computers based on the difference of their answers when classifying.
\end{definition}

From the perspective of the attack and defense game, adversarial examples and adversarial CAPTCHAs are different. In the case of adversarial examples, the deep learning models are on the defender side since adversarial examples’ main purpose is to attack the models. While in the case of adversarial CAPTCHAs, the deep learning models are on the attacker side since they try to recognize the CAPTCHAs. This indicates that we need to focus more on improving the security of the adversarial CAPTCHAs. Shi et al. stated that instead of generating human-imperceptible perturbations, generating human-tolerable perturbations makes adversarial CAPTCHAs more secure and easy to generate \cite{Shi2022}. The adversarial CAPTCHAs are more secure because the magnitude of the perturbations added to the original input is large, making the defense methods harder to eliminate the adversarial effect. Also, without restricting the perturbations to a small range, the perturbations that satisfied the requirement are easier to find.

\begin{figure*}[!t]
\centering
\includegraphics[width=1\linewidth]{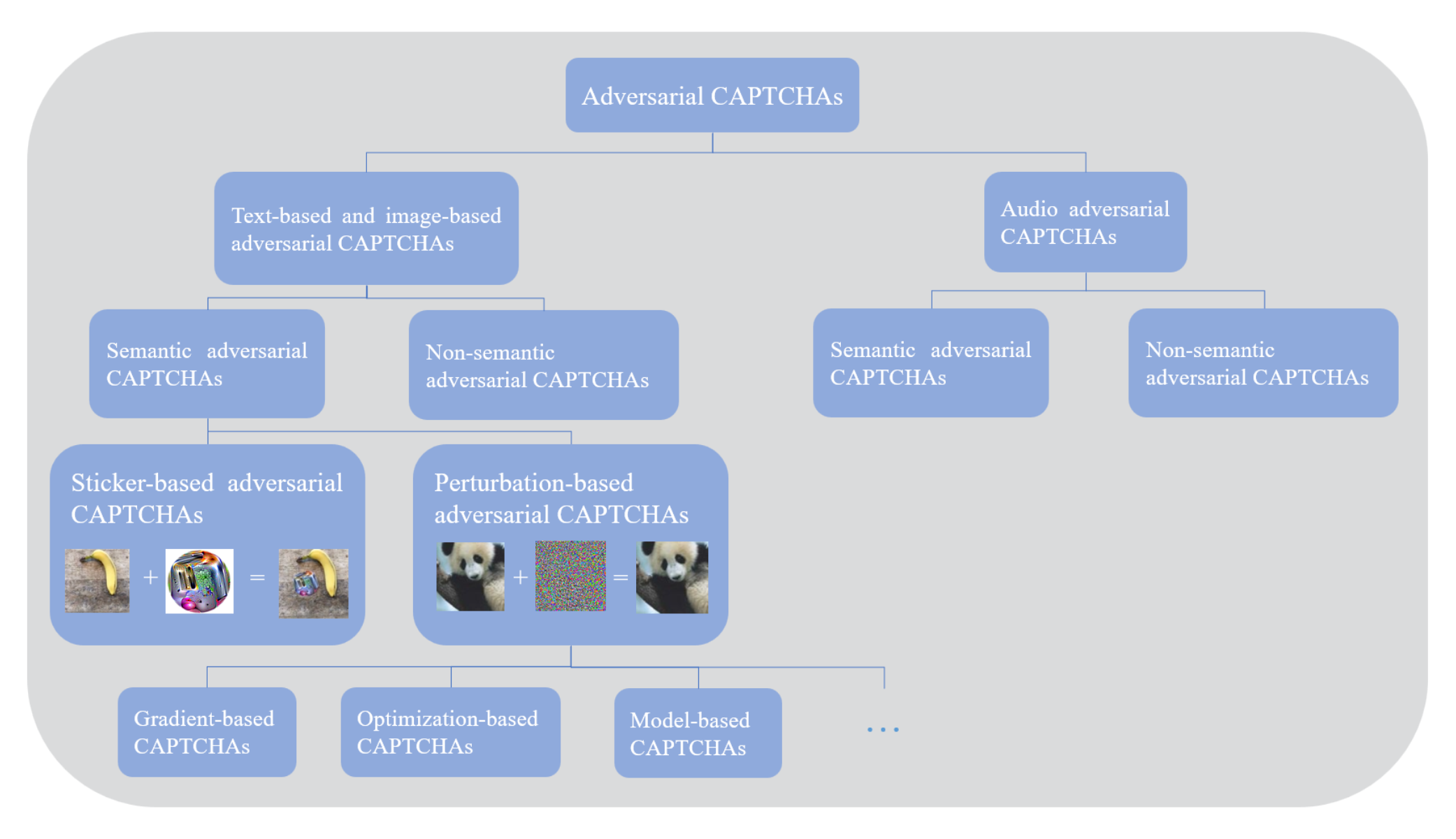}
\caption{Classification of adversarial CAPTCHAs}
\label{fig_7}
\end{figure*}

Figure \ref{fig_7} presents our way to categorize adversarial CAPTCHAs. Traditionally, adversarial CAPTCHAs refer to CAPTCHAs that applied a small magnitude of adversarial perturbations. Here, we extend the definition of adversarial CAPTCHAs to be more general, that is, any CAPTCHA that can fool the deep models. Based on the semantic information left in the adversarial CAPTCHAs, we categorize them into semantic adversarial CAPTCHAs and non-semantic adversarial CAPTCHAs: 

\subsection{Semantic adversarial CAPTCHAs}
Semantic adversarial CAPTCHAs are the most widely used adversarial CAPTCHAs. Affected by the adversarial perturbation in the CAPTCHA, DNN models will output a wrong answer, while humans can still give a correct answer, thus the computers and humans can be distinguished. In the semantic adversarial CAPTCHAs (recognizable CAPTCHAs), semantic information is still accessible to humans. For text-based and image-based adversarial CAPTCHAs, humans are able to recognize the characters or objects in the CAPTCHAs. And for audio adversarial CAPTCHAs, the audio is audible. For text-based and image-based CAPTCHAs, semantic adversarial CAPTCHAs can be further divided into 2 classes: adversarial perturbation-based CAPTCHAs and adversarial sticker-based CAPTCHAs.

The most common form of adversarial CAPTCHAs is constructed by adding adversarial perturbations, i.e., adding carefully crafted noise to the CAPTCHA. The perturbations added are usually imperceptible or quasi-imperceptible to humans, so the adversarial CAPTCHAs can only fool the computers. Brown et al. \cite{Brown2017} found another way to fool computers, which is by adding carefully crafted stickers (patches) to the CAPTCHA. The main disadvantage of the adversarial sticker is that the sticker is usually visible to humans. Compared to perturbation-based adversarial CAPTCHAs, sticker-based adversarial CAPTCHAs have the potential to implement adversarial attack in the physical world. In this case, the attacker side doesn't need to modify the input image since the attacker doesn't have access to the image, on the contrary, the attacker modifies the objects captured by the camera or the camera lens in the physical world. Several attempts have been made and testified in real applications, such as the adversarial glasses \cite{Sharif2016}, adversarial camera stickers \cite{Li2019}, adversarial stickers \cite{Wei2022} and AdvHat \cite{Komkov2021}.

\subsection{Non-semantic adversarial CAPTCHAs}
In the non-semantic adversarial CAPTCHAs, humans cannot recognize the input since there is no semantic information, while the DNNs can somehow recognize the input \cite{Nguyen2015}. Because semantic information is not included in the non-semantic adversarial CAPTCHA, humans are not able to recognize the content in the CAPTCHA. On the contrary, DNN models can recognize the CAPTCHA and output a specific answer, thus the CAPTCHA system can distinguish between humans and computers.
In the following sections, we will introduce some general methods to generate adversarial CAPTCHAs and some examples that have been successfully applied to adversarial CAPTCHAs.

% --------------------------Text-based and image-based adversarial CAPTCHAs---------------------------

\section{Text-based and image-based adversarial CAPTCHAs}
Here we slightly introduce the progress of text-based CAPTCHAs and image-based CAPTCHAs verification. Both text-based and image-based CAPTCHAs are presented in image form. Text-based CAPTCHAs require users to recognize the characters in the image and type the corresponding characters. Image-based CAPTCHAs usually require users to recognize the object in the image and select the corresponding name or category of the object.

Many approaches are introduced to generate an adversarial example of an image. Since text-based and image-based CAPTCHAs are presented in a form of an image, we introduced the text-based and image-based adversarial CAPTCHAs together. The difference between generating adversarial examples for text-based and image-based CAPTCHAs is that the magnitude of the perturbation added to the text-based CAPTCHAs is larger than image-based CAPTCHAs since the content in the text-based CAPTCHAs is simpler than image-based CAPTCHAs. The summary of adversarial CAPTCHAs is listed in TABLE I.

Generating adversarial CAPTCHA is transforming the original image presented to the users into an adversarial example. Unlike the requirement of adversarial examples that the perturbations should be imperceptible, the perturbations of adversarial CAPTCHA can be flexible. This means that the perturbations can be perceptible to humans, as long as they won’t affect the usability of CAPTCHA. Here we listed some classic approaches to generate adversarial examples and some methods that applied successfully to CAPTCHAs.

\subsection{Gradient-based methods}
Since loss value is used to predict the classification result, it is natural to utilize gradient information to alter the loss value, thus changing the final prediction. The essence of generating an adversarial example using gradient-based methods is to find the local maxima distributed on the loss surface. Here are some attack methods based on gradient

\subsubsection{Fast Gradient Sign Method}
\begin{enumerate}[a)]
    \item{Fast Gradient Sign Method, or FGSM in abbreviation, is first proposed by Goodfellow et al. in 2015 \cite{Goodfellow2014}, it is one of the classic approaches to generate an adversarial example. Intuitively, increasing the model’s loss of an image can make the model misclassify. Following this idea, Goodfellow et al. change the pixel’s intensity of an image by one large step, making its loss increase while ensuring the alteration is small enough. FGSM is simple yet effective. Many efforts were made to improve FGSM, below are some variants we’d like to introduce.}
    \item{Iterative-FGSM, or I-FGSM in abbreviation, is proposed by Kurakin et al. \cite{Kurakin2016}. FGSM only takes one large step toward the direction that increases the loss value, the perturbations may not be robust enough to fool the model. I-FGSM is an improved method based on FGSM, it takes multiple small steps toward the direction that increases the loss value. Given the fact that the grayscale for a pixel ranges from 0 to 255, I-FGSM also clips the value of a pixel when it is out of range during the process of generating adversarial examples. Intuitively, FGSM might not be stepping close enough to the local maxima by only stepping one step. However, I-FGSM could steps closer through multiple iterations, creating more robust perturbations and only slightly increasing the generation time.}
    \item{Projected Gradient Descent, or PGD in the abbreviation is known as the most powerful first-order attack \cite{Madry2017}. Compared to I-FGSM, PGD increases iterations and starts with random noise. There are lots of local maxima on the loss surface, I-FGSM might not find a better local maximum due to its initial location. By introducing random noise to start in random locations, PGD could find a better local maximum theoretically.}
    \item{Momentum iterative-FGSM, or MI-FGSM in abbreviation, is an extension to I-FGSM incorporating momentum \cite{Dong2018}. Similar to PGD, MI-FGSM also attempts to find a better local maximum by introducing a momentum term, making it not stranded in a specific local maximum, thus generating a more powerful adversarial example. However, its performance is also restricted by its initial location.}
\end{enumerate}

\subsubsection{Jacobian-based Saliency Map Attack}
A saliency map is a matrix that indicates the importance of a pixel for a model to predict the corresponding class. According to the property of the saliency map, it is obvious that changing the most important pixel of an image might have a great chance to make the model misclassify. Inspired by the saliency map in the computer vision domain, Papernot et al. proposed Jacobian-based Saliency Map Attack (JSMA) \cite{Papernot2016}. JSMA computes the saliency map utilizing the logits of a model given an input and alters the most important pixel base on the saliency map.

\subsubsection{DeepFool}
Moosavi-Dezfooli et al. proposed the DeepFool algorithm \cite{Moosavi2016}. In the classification problem, any image can be represented by a point in high-dimensional space, the classification results are separated by hyperplanes (decision boundaries) in high-dimensional space. At each iteration, the DeepFool algorithm leverages gradient to compute the distances to each decision boundary, then perturb the image by a small vector to push the image towards the nearest incorrect decision boundary and thus fool the model. Compared to FGSM, DeepFool can fool the models with smaller perturbations.

\subsection{Optimization-based methods}
\subsubsection{L-BFGS}
Limited-memory BGFS for the adversarial attack, or L-BFGS in abbreviation, is proposed by Szegedy et al. \cite{Szegedy2013}. The problem is formalized by finding the perturbation that makes the model misclassify and the perturbation remains as minimum as possible. While theoretically, the adversarial examples generated by this method are powerful, the computational overhead for solving this problem remains high.

\subsubsection{Carlini \& Wagner Attack}
Carlini and Wagner proposed CW attack based on optimization \cite{Carlini2016}. Similar to L-BGFS, the CW attack also attempts to find a perturbation that fools the model while the magnitude of the perturbation remains minimum. Different from L-BGFS, the CW attack uses the gradient of the logits instead of the gradient of SoftMax. And the authors use 3 different distance metrics, $L_0$, $L_2$, and $L_{\infty}$ respectively. By $L_0$ distance, the number of altered pixels is constrained to minimal, and the $L_2$ distance minimizes the Euclidean distance between the original image and the adversarial example. And by $L_{\infty}$ distance, the magnitude of the pixel with maximal alteration is constrained to a certain level. When adding adversarial perturbation to the benign image, the pixel value might exceed its valid range. Most methods handle this situation by clipping the pixel value. However, such a method might affect the attack performance. CW attack handles this situation by introducing the $tanh$ function to the perturbation, which makes the pixel value always valid. Although the CW attack can generate optimal adversarial examples theoretically, it can take more than a hundred seconds to generate an adversarial example given the fact that solving this problem is complicated, while methods like FGSM only take less than 0.1s.

\subsubsection{Universal Perturbations}
Unlike other methods that generate specific perturbations for given images, Moosavi-Dezfooli et al. proposed a universal method, i.e., generate a universal perturbation that can fool the model when added to any images \cite{Moosavi2017}. In high-dimension space, each image is represented as a point in the space, and what determines the classification result are various regions enclosed by the decision boundaries. Suppose there are various data points in the high-dimension space, and the algorithm tries to find a vector that pushes images outside the decision boundary of the first data point. Then the algorithm tries to find a vector that when added to the first vector, the new vector can push images outside the first and second decision boundaries. By repeating the same method, a universal perturbation that fools the model against any image can be found.

\subsection{Evolutionary algorithm methods}
Su et al. proposed one-pixel attack \cite{Su2019}, namely generating adversarial examples with only one pixel changed. One pixel attack is based on the differential evolution algorithm, starting from a randomized population (possible solutions). At each iteration, every solution mutates and becomes a different solution based on different policies. After the mutation process, different mutations will be mixed to create a mixed mutation. Then the performance of all solutions to fool the model will be evaluated and only the plausible solutions will be reserved and repeat the process above.

\subsection{Generation model methods}

\subsubsection{Adversarial Transformation Networks}
Baluja and Fischer et al. proposed adversarial transformation network \cite{Baluja2017}, which utilized multiple feed-forward networks to generate adversarial examples. The adversarial examples are generated by restricting the loss function. The loss function consists of two parts, the first part limits the magnitude of perturbations to make the perturbations imperceptible, and the second part aims to make the model give the wrong prediction.

\subsubsection{AdvGAN}
Similar to ATN, Xiao et al. proposed AdvGAN which uses a GAN to generate adversarial examples \cite{Xiao2018}. Specifically, the generator and discriminator in GAN are both encoder-decoder structures. After finishing the training process of AdvGAN, it can generate a new adversarial example without accessing the target network’s loss function, so the adversarial example generation time is less than the iterative methods such as I-FGSM. Jandial et al. proposed AdvGAN++, an improved version of AdvGAN \cite{Jandial2019}. Using the vulnerability of the latent layer of the model \cite{Kumari2019}, AdvGAN++ can generate more robust adversarial examples. Compared to ATN, adversarial examples generated by GAN are more realistic.

\subsection{Methods that successfully applied to text-based and image-based CAPTCHAs}

\subsubsection{aCAPTCHA}
Based on JSMA, Shi et al. proposed JSMA-f, where f stands for frequency domain \cite{Shi2022}. The authors convert images from the space domain to the frequency domain and perform JSMA to the high-frequency components (HFC) of the frequency domain. The perturbations generated by JSMA-f are a global change to the image, which is more difficult to remove compared to local changes in the space domain. Besides, the JSMA-f algorithm alters the neighborhood pixels as well because of their partial similarity to the candidate pixel and its neighborhood pixels, making it generates adversarial CAPTCHAs faster. The same procedure is also introduced to CW attack to generate adversarial text-based CAPTCHAs. For image-based adversarial CAPTCHAs, the authors compute the adversarial perturbation in an iterative way similar to I-FGSM.

\subsubsection{Immutable Adversarial Noise}
Osadchy et al. proposed immutable adversarial noise (IAN) that is specifically resistant to filtering defense (will be introduced in section V) \cite{Osadchy2017}. The authors experimented and discovered that the median filter is the most capable one to filter adversarial noise. Similar to I-FGSM, they generate adversarial examples with multiple iterations. At each iteration, they check if the current generated adversarial examples are recognizable by DNNs after median filtering. If the current generated image is breakable, they continue the generation process, until the adversarial noise is no longer filtered by the median filter. Then they proposed the DeepCAPTCHA scheme. It presents an image with IAN to the user and requires the user to select images that are in the same class as the presented image.

\subsubsection{Robust text CAPTCHA}
Shao et al. proposed a scaled Gaussian translation with channel shift attack for adversarial text CAPTCHA \cite{Shao2021}. The authors introduced scaling, spatial translation and channel shift to improve the diversity of the images, thus improving the transferability of adversarial examples generated according to a previous work by Xie et al. \cite{Xie2019}. Furthermore, the authors utilized Nesterov Accelerated Gradient (NAG) to achieve better performance. To reduce computational complexity, they assumed the gradient remains the same after conducting a small parallel shift and scaling.

\subsubsection{CAPTURE}
Hitaj et al. proposed a new CAPTCHA scheme named CAPtcha Technique Uniquely Resistant (CAPTURE) \cite{Hitaj2021}. CAPTURE is a CAPTCHA scheme that combined semantic and non-semantic adversarial CAPTCHAs. The authors involved adversarial patches and unrecognizable images in the CAPTCHA system. A representative work of unrecognizable images is proposed by Nguyen \cite{Nguyen2015}, who generated unrecognizable images using evolutionary algorithms and gradient descent. The adversarial patch is a sticker that when applied to an image, has an adversarial effect while being independent of the image. The authors utilized these approaches to CAPTURE. CAPTURE presents a few images to the users and requires the user to select images that are real images of a certain object.

\subsubsection{AdvCAPTCHA}
Shi et al. proposed an adversarial CAPTCHA generation system advCAPTCHA \cite{Shi2020}. In a real-world scenario, the details of the attacker model are usually unknown. It requires us to generate adversarial CAPTCHAs in a black-box setting. The authors trained a Convolutional Recurrent Neural Network (CRNN) to substitute for the unknown CAPTCHA solver model. They generated adversarial CAPTCHAs by decreasing the CTC \cite{Graves2006} of the substitute model and utilized a mask matrix to constrain the position where perturbations were added. To better imitate the attacker model, they distributed the adversarial CAPTCHAs generated and recorded the feedback from the attacker model, then they fine-tuned the substitute model with these collected data. After fine-tuning, the authors generated new adversarial CAPTCHAs and repeated the same process for certain times.

\subsubsection{Involving spatial smoothing to adversarial CAPTCHAs}
Spatial smoothing is an image processing technique that can be used as a defense method to mitigate adversarial perturbations, but there is a trade-off in spatial smoothing. While being able to remove adversarial perturbations, excessive spatial smoothing can blur the image and corrupt the semantic information in it. Matsuura et al. proposed a method that applies spatial smoothing before adding adversarial perturbations generated by FGSM to the original image \cite{Matsuura2021}. When the attacker wants to remove adversarial perturbations using spatial smoothing, excessive spatial smoothing corrupts the semantic information and the attacker fails to recognize the characters. However, the adversarial CAPTCHAs they generated have relatively poor usability for users to recognize.

\subsubsection{Adversarial CAPTCHAs with low-frequency perturbations}
As mentioned before, Osadchy et al. stated that the median filter can efficiently remove the adversarial \cite{Osadchy2017}. Terada et al. mitigated this disadvantage by applying adversarial perturbations in low-frequency components, where the median filter cannot efficiently remove the perturbations \cite{Terada2022}. Similar to IAN generation, they computed the perturbations iteratively until the median filter failed to remove the adversarial effect.

\subsubsection{Adversarial CAPTCHA scheme utilizing user behavior}
Similar to reCAPTCHA, Zheng et al. proposed a CAPTCHA scheme that utilizes users’ behavior to defend against bots \cite{Zheng2021}. The CAPTCHA scheme first presents a benign CAPTCHA to the user, based on the user’s recognition time and accuracy, the CAPTCHA system judges if the user is a bot. If the system cannot determine, it will present an adversarial CAPTCHA generated by FGSM to the user. If the user cannot recognize the CAPTCHA presented, it will be judged as a computer program.

\begin{table*}[]
\caption{Summary of text-based and image-based adversarial CAPTCHAs
\label{tab:table1}}
\centering
\begin{adjustbox}{max width=1.1\linewidth,center}
\begin{tabular}{ccccccc}
\hline
Authors         & Category                      & Attack Type                                                   & Year & Threat Models                                                                                & Dataset                                                     & Examples     \\ \hline
Shi et al. \cite{Shi2022}     & Perturbation-based            & White-box                                                     & 2021 & \begin{tabular}[c]{@{}c@{}}NetInNet\\ GoogleNet\\ VGG\\ ResNet50\end{tabular}                & \begin{tabular}[c]{@{}c@{}}MNIST\\ ILSVRC-2012\end{tabular} &  \begin{tabular}[c]{@{}c@{}}\includegraphics[width=0.3\textwidth]{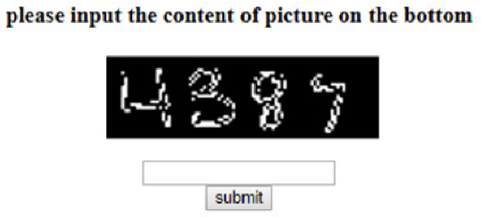}\\ \includegraphics[width=0.3\textwidth]{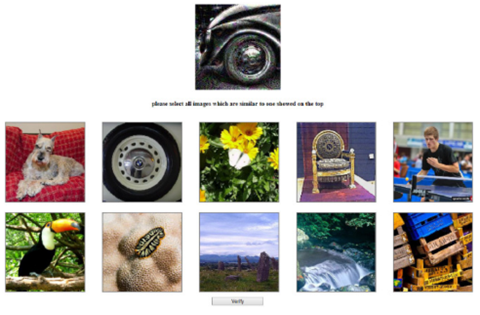}\end{tabular} \\ \hline
Osadchy et al. \cite{Osadchy2017}  & Perturbation-based            & White-box                                                     & 2017 & \begin{tabular}[c]{@{}c@{}}CNN\\ AlexNet\end{tabular}                                        & \begin{tabular}[c]{@{}c@{}}MNIST\\ ILSVRC-2012\end{tabular} &  \begin{tabular}[c]{@{}c@{}}\includegraphics[width=0.3\textwidth]{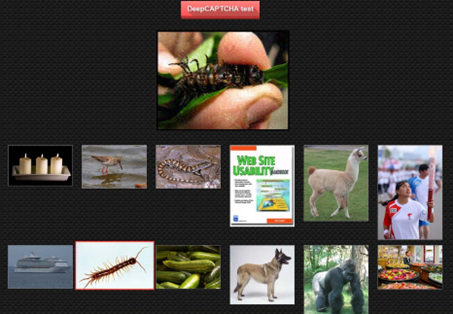} \end{tabular}   \\ \hline
Shao et al. \cite{Shao2021}    & Perturbation-based            & \begin{tabular}[c]{@{}c@{}}White-box\\ Black-box\end{tabular} & 2021 & \begin{tabular}[c]{@{}c@{}}LeNet\\ AlexNet\\ VGG\\ GooLeNet\\ ResNet\\ DenseNet\end{tabular} & \begin{tabular}[c]{@{}c@{}}MNIST\\ EMNIST\end{tabular}      &   \begin{tabular}[c]{@{}c@{}}\includegraphics[width=0.3\textwidth]{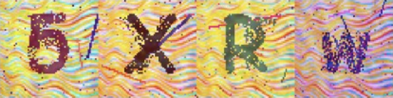} \end{tabular} \\ \hline
Hitaj et al. \cite{Hitaj2021}   & Sticker-based \& Non-semantic & White-box                                                     & 2021 & \begin{tabular}[c]{@{}c@{}}InceptionV3\\ Xception\\ VGG\\ ResNet\\ MobileNet\end{tabular}    & ImageNet                                                    &    \begin{tabular}[c]{@{}c@{}}\includegraphics[width=0.3\textwidth]{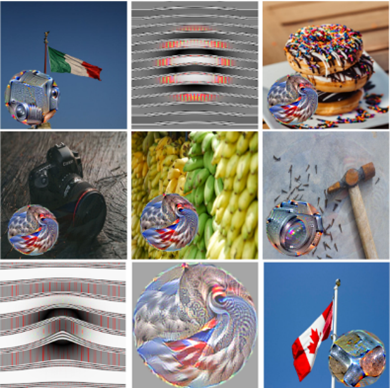} \end{tabular}     \\ \hline
Shi et al.  \cite{Shi2020}    & Perturbation-based            & Black-box                                                     & 2020 & CRNN (substitute model)                                                                      & Self-collected dataset                                      & Not provided \\ \hline
Matsuura et al. \cite{Matsuura2021} & Perturbation-based            & White-box                                                     & 2021 & CNN                                                                                          & Self-collected dataset                                      &   \begin{tabular}[c]{@{}c@{}}\includegraphics[width=0.3\textwidth]{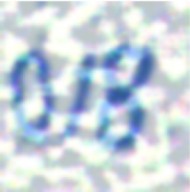} \end{tabular}   \\ \hline
Terada et al. \cite{Terada2022}  & Perturbation-based            & White-box                                                     & 2022 & \begin{tabular}[c]{@{}c@{}}MLP\\ VGG\end{tabular}                                            & \begin{tabular}[c]{@{}c@{}}MNIST\\ Caltech-256\end{tabular} &  \begin{tabular}[c]{@{}c@{}}\includegraphics[width=0.3\textwidth]{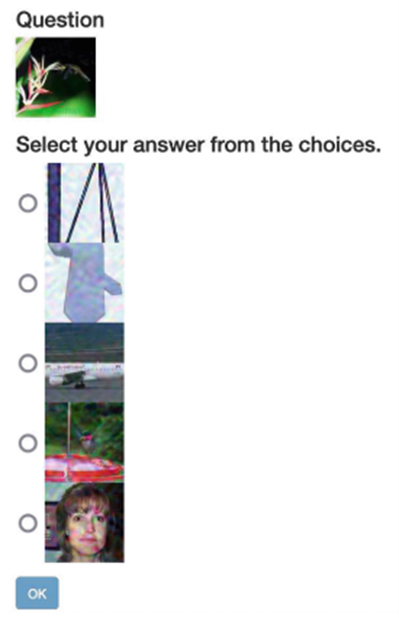} \end{tabular}  \\ \hline
Zheng et al.  \cite{Zheng2021}  & Perturbation-based            & White-box                                                     & 2021 & \begin{tabular}[c]{@{}c@{}}CNN\\ Xception\end{tabular}                                       & Self-collected dataset                                      & Not provided \\ \hline
\end{tabular}
\end{adjustbox}
\end{table*}

% -----------------------------------Audio adversarial CAPTCHAs--------------------------------------

\section{Audio adversarial CAPTCHAs}
The existence of visual-impaired users requires CAPTCHA in audio form. Audio CAPTCHA presents an audio sample to the users and requires the users to recognize the content in the audio sample. The main problem of audio CAPTCHA is that the content in the audio can be recognized by various voice recognition models. A common approach to avoid this is to add random noise to the audio, but there is a trade-off to this approach, random noise with a small magnitude can still be recognized by deep models and excessive noise will affect the usability of CAPTCHA. Thus, audio adversarial CAPTCHA is needed. We aim to find specific noise that can fool the deep models while maintaining the good usability of CAPTCHAs. 

Compared to visual recognition models, audio recognition models require preprocessing. One of the most common approaches for audio preprocessing is Mel-frequency Cepstrum (MFC) transformation. Instead of using the frequency scale measured in hertz, MFC utilizes the Mel scale. Given the fact that human ears percept differently to the same difference in high-frequency sound and low-frequency sound, a normal scale measured with hertz cannot represent the way humans percept the sound. Thus, the Mel scale is introduced to imitate how humans percept sound. The same difference in the Mel scale allows humans to feel the same difference in pitch. Another reason for choosing MFC transform is human ears only percept sound in certain frequencies. Mel scale utilizes this special property of human ears, it filters the useless frequencies and only reserved the frequencies that are perceptible by a human.

The overall process of MFC transformation can be summarized as follows:
\begin{enumerate}[Step 1.]
    \item{The input audio is transformed into the frequency domain using discrete Fourier transform (DFT)}
    \item{The frequency scale converts from hertz to Mel}
    \item{Triangular filter is applied to simulate the human auditory system}
    \item{Computing Mel-frequency cepstrum by applying logarithm of powers}
    \item{Mel-frequency cepstrum coefficients are computed by applying DFT}
\end{enumerate}

Unlike models for image recognition, in which all the layers are differentiable, models for audio recognition take inputs after heavy preprocessing and feature extraction (e.g., MFCC transform). The triangular filter quantifies certain frequencies of audio into a specific Mel frequency, which sets an obstacle to the backpropagation through the MFC layer, thus generating adversarial audio examples is challenging. We introduce some methods to generate audio adversarial examples in the following content.

\subsection{Parameter-tuning methods}
\subsubsection{Cocaine Noodles}
Vaidya et al. proposed Cocaine noodles to generate sounds with an adversarial effect \cite{Vaidya2015}. Voice command is given as an input to the audio mangler, where the audio mangler modifies the original input by adjusting the MFCC parameters and extracts the corresponding features. After extraction, the audio signal is generated by performing a reverse MFCC. Having the adversarial effect though, audio signals generated by cocaine noodles are highly recognizable and can easily be detected by human ears.
    
\subsubsection{Hidden voice commands}
Mishra et al. proposed hidden voice commands, which is an extensive work of cocaine noodles \cite{Carlini2016H}. Compared to cocaine noodles, the audio signal generated by hidden voice commands is difficult for a human to recognize. Hidden voice commands can be seen as an iterative version of cocaine noodles, the generation process is repeated while the audio signal is still recognizable by a human.

\subsection{Optimization-based methods}
\subsubsection{CW attack on audio}
Carlini and Wagner generated audio adversarial examples for transcription models by applying CW attack to audio \cite{Carlini2018}. The magnitude of perturbation is measured in decibels. The authors formalize the problem in which the transcription generated by the model is perturbed while the decibel of the perturbation is minimal. However, audio transcription is not a classification problem, the data are pairs of audio and its corresponding transcription text. The texts are of variable length and there is no alignment between the audio and the text, solving the optimization problem directly is difficult. Therefore, the authors used Connectionist Temporal Classification loss (CTC loss) proposed by Graves et al. to solve this problem \cite{Graves2006}. The CTC loss is a differentiable metric to measure the distance between the model output and the true target phrase. Thus, the optimization problem turned into finding the perturbation that minimizes the CTC loss while the perturbation remains minimal. The limitation of CW attack is that the perturbation must be intact when performing an attack, thus it cannot be played over the air like hidden voice commands and the dolphin attack.

\subsubsection{Imperceptible, robust and targeted audio adversarial examples}
Qin et al. proposed a method to generate audio adversarial examples which were imperceptible and robust enough to be played over-the-air \cite{Qin2019}. To achieve imperceptibility, the authors utilized frequency masking, which refers to the phenomenon that a louder signal can make other signals at nearby frequencies imperceptible. The generation process is separated into 2 stages: the first stage focused on fooling the DNN model and the second stage made the perturbation imperceptible. The experiment showed that the proposed method attacked the Lingvo model with 100\% attack success rate, and the attack success rate was above 50\% even when played over the air.

\subsection{Gradient-based methods}
\subsubsection{Houdini attack}
Tasks like human pose estimation, semantic segmentation, and speech recognition are combinatorial and non-decomposable, where gradient descent does not apply to these tasks. However, one can replace the task loss with a differentiable surrogate loss. Cisse et al. proposed the Houdini algorithm that is tailored to task losses to create adversarial examples \cite{Cisse2017}. For the speech recognition task, the authors utilized components in the CTC-loss to generate surrogate loss and successfully attacked an ASR (Automatic Speech Recognition) system.

\subsubsection{Attack via psychoacoustic hiding}
Schonherr et al. implemented an attack via psychoacoustic hiding. Psychoacoustic hearing thresholds describe the relation between human auditory perception and sound frequencies. Some parts of the input audio signal may not be heard by humans, thus providing chances to modify the audio signals without alerting humans. The authors combined the preprocessing step and the DNN recognition step into a joint DNN, allowing them to update the raw audio directly through the backpropagation of the joint DNN. The psychoacoustic hearing thresholds were calculated using the raw audio, and the gradient of the magnitude of the spectrum was scaled by the thresholds.

\subsection{Hardware-targeted methods}
Zhang et al. proposed the dolphin attack, which integrates voice commands in ultrasonic frequency \cite{Zhang2017}. Given the fact that humans cannot hear ultrasound, thus dolphin attack achieved inaudibility. The dolphin attack utilizes the nonlinearity of the Micro Electro Mechanical Systems (MEMS) microphones and the amplifier to generate a high-frequency sound, which can be demodulated to recognizable voice commands by the Voice Controllable Systems (VCS). Experiments show that voice assistants like Alexa and Siri can respond to these inaudible voice commands. However, such an attack may fail when the high-frequency sound is filtered.

\subsection{Genetic algorithm methods}
Alzantot et al. proposed a genetic method to generate audio adversarial examples \cite{Alzantot2018}. The algorithm first generates candidates of adversarial examples by adding random noise to a subset of samples. Candidates that meet the requirements become part of the next generation. Members in the same generation can be mixed to generate a new candidate, and a candidate can mutate into a new candidate by adding random noise. This process iterates until the attack is successfully done.

\subsection{Methods that successfully applied to audio adversarial CAPTCHAs}
\subsubsection{aaeCAPTCHA}
Hossen and Hei designed and implemented an audio adversarial CAPTCHA (aaeCAPTCHA) system \cite{Hossen2022}. They implemented FGSM and PGD attack on audio. Because the MFCC transformation is indifferentiable, they used a differentiable implementation of MFCC by Carlini et al. \cite{Carlini2018} instead. The aaeCAPTCHA scheme requires the user to transcribe the adversarial audio to prove they are humans. For usability evaluation, the results show that audio within 8 words can achieve good usability.

\subsubsection{Utilizing I-FGSM and DeepFool in audio adversarial CAPTCHA generation}
Shekhar et al. generated audio adversarial CAPTCHA using I-FGSM and DeepFool algorithms \cite{Shekhar2019}. They evaluated the attack success rate (ASR) on traditional machine learning models and DNNs, and achieved initial success. They also utilized some of the audio adversarial examples with different magnitudes of perturbations generated by I-FGSM and DeepFool algorithms to retrain the attack model. After retraining, the accuracy of the attack model still remained around 25-36\%. The audio adversarial CAPTCHAs are still effective as long as the attacker has less than 46\% of the sample from different perturbation levels of audio adversarial examples.

\begin{table*}[]
\caption{Summary of audio adversarial CAPTCHAs\label{tab:table2}}
\centering
\begin{adjustbox}{max width=1\linewidth,center}
\begin{tabular}{cccccc}
\hline
Authors        & Category       & Attack Type & Year & Threat Models                                                                                                & Dataset                \\ \hline
Hoosen and Hei \cite{Hossen2022} & Gradient-based & White-box   & 2022 & \begin{tabular}[c]{@{}c@{}}Deep Speech\\ Deep Speech 2\\ Jasper\\ Wave2Letter+\\ Lingvo\\ Kaldi\end{tabular} & LibriSpeech            \\ \hline
Shekhar et al. \cite{Shekhar2019} & Gradient-based & White-box   & 2019 & \begin{tabular}[c]{@{}c@{}}CNN\\ VGG\end{tabular}                                                            & Self-generated dataset \\ \hline
\end{tabular}
\end{adjustbox}
\end{table*}

% ---------------------------------Challenges of adversarial CAPTCHAs--------------------------------

\section{Challenges of adversarial CAPTCHAs}
Although adversarial CAPTCHAs can fool deep models and disable recognition, adversaries can take countermeasures to transform adversarial examples into benign examples. This requires adversarial CAPTCHAs to have the robustness to break these defenses. Here are several approaches to achieve defenses against adversarial CAPTCHAs.

\subsection{Adversarial Training}
Proposed by Goodfellow et al., Adversarial training is a technique using fine-tuning to make the model adapt to adversarial examples \cite{Goodfellow2014}. The adversarial examples are annotated with correct labels and fine-tuned the model with these data. While the accuracy for adversarial examples increased after adversarial training, the accuracy for benign examples might slightly decrease. Also, because adversarial examples can fool the deep models, they require humans to relabel the images, which can bring a high cost.

\subsection{Image Filtering}
Filtering is a defense method only suitable for image adversarial examples. The idea of filtering is that the perturbation of an adversarial example can be seen as some noise added to the images, filtering might be able to denoise the images and eliminate their adversarial effect. Common filters like median filter and Gaussian filter can be used to defend adversarial examples. However, there is a trade-off between defense ability and accuracy. A bigger kernel size can achieve strong defense ability, while at the same time the semantics of the image might be corrupted, making humans or deep models unable to recognize it.

\subsection{Defensive Distillation}
Model distillation proposed by Hinton et al. is a technique to reduce model complexity by transferring knowledge from a sophisticated model to a smaller model \cite{Hinton2015}, so as to reduce the computational overhead. Based on model distillation, Papernot et al. proposed defensive distillation, a method applicable to any network to improve robustness by feeding the probability vector predictions to the distilled network \cite{Papernot2016D}. However, Carlini and Wagner et al. claimed that their proposed method of CW attack successfully attacked the distilled network \cite{Carlini2016}.

\subsection{Gradient Masking}
Some of the attack methods like FGSM utilize gradient to perform attacks. Therefore, making the model provide useless gradients can fail the attack. After performing gradient masking, the loss surface of the model is distorted and makes it hard for an adversary to find an adversarial example near the decision boundary. However, this defense method is only applicable to gradient-based attack methods, methods without utilizing gradients like one-pixel attack cannot be defended.

\subsection{Incremental Learning}
Na et al. proposed an incremental learning method to defend against adversarial CAPTCHAs \cite{Na2020}. The authors utilized a deep model to recognize CAPTCHAs. The CAPTCHA system returns feedback after the deep model submitted the answers. Based on the feedback, if the model submitted a wrong answer, several recent adversarial CAPTCHAs will be manually labeled and use these data to fine-tune the model. Through this constant learning process, the robustness of the deep model improves, but manual labeling of the data still requires human effort with a high cost.

Another challenge for adversarial CAPTCHA is generating adversarial examples in a black-box setting. Most adversarial example generation methods are in white-box or semi-white box settings, in which all the details of the model including model structure, and loss function are known. However, in most cases in the physical world, an adversary has no access to the original model, only the inputs and their corresponding outputs are known. Two common ways to implement a black-box attack are model distillation and utilizing the transferability of adversarial examples. Model distillation aims to imitate the threat model by feeding the inputs and the model’s corresponding outputs to a new model, then attack the new model utilizing various attack methods since the details of the model are known. The transferability of adversarial examples is proposed by Szegedy et al., where adversarial examples for a deep model may also attack another deep model successfully.

% ---------------------------------Possible direction for future research----------------------------

\section{Possible direction for future research}
As stated in the last section, the challenges of adversarial CAPTCHAs are defense methods and black-box settings, thus the directions for future research are specifically tied to these challenges. The most common directions of research to mitigate these challenges are constructing a surrogate model, and utilizing the transferability of adversarial examples.

\subsection{Constructing surrogate models}
To implement black-box attack, one way is to imitate the black-box model and generate a surrogate model, then conduct a white-box attack on the surrogate model since the details of the model are fully accessible. The most common methods utilized model distillation proposed by Hinton et al. to distill a surrogate model \cite{Hinton2015}. Though the distilled model might act in the same way as the black-box model when fed with benign data, it might act differently when fed with adversarial examples. To mitigate this situation, Xiao et al. proposed dynamic distillation that further fine-tunes the surrogate model with adversarial examples \cite{Xiao2018}.

\subsection{Improving transferability}
The transferability of an adversarial example is a phenomenon that adversarial examples that successfully attacked one model can successfully attack another model with high probability. The underlying mechanism of transferability is still unknown, but Wu et al. proposed a possible explanation: though diverse models are different in their model structures, they share a similar attention region on the same image. When adversarial perturbation can affect the original attention region of one model, chances are high that it can affect other models \cite{Wu2020}. Many researchers tried to improve the transferability of adversarial examples with different methods and this direction is still a popular one for implementing black-box attack \cite{Wu2020,Huang2022,Byun2022,Dong2022,Jang2022,Liu2022}.

\subsection{Unexploited method to generate adversarial examples}
Most common methods to generate adversarial examples are based on explicit algorithms, thus coming up with an algorithm that achieves higher resistance to various defense methods could be extremely difficult.

A few attempts have been made to utilize the learning capacity of deep learning models, like ATN and AdvGAN we introduced before in section III. 

However, to our knowledge, reinforcement learning hasn’t been widely introduced to generate adversarial examples. The most classic algorithm of reinforcement learning is the Q-learning algorithm. Traditional Q-learning algorithm requires storing a Q-table to take a specific action, where the Q-table consists of the expected return the agent will receive if it starts in a given state and takes a specific action. Consider using reinforcement learning to generate adversarial examples, the states are denoted by a specific image. In this case, the size of the state space is close to infinity, thus it is impossible to store a Q-table with unlimited size. Deep reinforcement learning substitutes the Q-table with a deep learning model. The deep model takes a state as input and outputs the actions and their corresponding Q-value without storing every state. Such a method is called Deep Q-Network (DQN). Xiao et al. utilize DQN to generate adversarial examples on the MNIST dataset and achieved initial success \cite{Xiao2020}.

To its nature, deep reinforcement learning is suitable for black-box attacks theoretically as one can set the reward by only comparing the probability vector output by the target model concerning the original image and the generated image. By giving a penalty when the generated images failed to build resistance to the defense methods, the deep reinforcement learning agent may find a way to generate adversarial examples that are resistant to various defense methods. However, what we’ve discussed by now is just a theoretical analysis, the actual effectiveness of utilizing deep reinforcement learning remains to be explored.

% ---------------------------------------------Conclusion-------------------------------------------

\section{Conclusion}
 In this paper, we refined the definition of adversarial CAPTCHAs and systematically summarized current research related to adversarial CAPTCHAs. We first presented the history and some basic concepts of CAPTCHA. Adversarial CAPTCHAs can mitigate the trade-off between security and usability in CAPTCHAs. Then we reviewed some of the methods used to generate adversarial CAPTCHAs. However, as a method to attack the adversaries' CAPTCHA recognition models, some corresponding defense methods can be used to counter adversarial CAPTCHAs. We reviewed some general methods used to defend adversarial CAPTCHAs and finally discussed some possible directions for further research on adversarial CAPTCHAs.

%%
%% The acknowledgments section is defined using the "acks" environment
%% (and NOT an unnumbered section). This ensures the proper
%% identification of the section in the article metadata, and the
%% consistent spelling of the heading.
\begin{acks}
This work is supported by the National Natural Science Foundation of China (61976142) and Shenzhen Science and Technology Plan Project (JCYJ20210324093609025).
\end{acks}

%%
%% The next two lines define the bibliography style to be used, and
%% the bibliography file.
\bibliographystyle{ACM-Reference-Format}
\bibliography{export}

\end{document}